\documentclass[12pt]{iopart}
\usepackage{graphicx}% Include figure files
\usepackage{dcolumn}% Align table columns on decimal point
\usepackage{bm}% bold math
\usepackage[mathscr]{eucal}
\usepackage{mathrsfs}
\usepackage{cite}

\usepackage{txfonts}

\begin{document}

% Full title of the paper (Capitalized)
\title[Overestimation of melting temperature]{Overestimation of melting temperatures calculated by first-principles molecular dynamics simulations}

% Authors, for the paper (add full first names)
\author{Koun Shirai$^{1,2}$, Hiroyoshi Momida$^3$, Kazunori Sato$^4$, and Sangil Hyun$^{5}$}

\address{
$^1$Vietnam Japan University, Vietnam National University, Hanoi, \\
Luu Huu Phuoc Road, My Dinh 1 Ward, Nam Tu Liem District, Hanoi, Vietnam \\
$^2$ SANKEN, Osaka University, 8-1 Mihogaoka, Ibaraki, Osaka 567-0047, Japan \\

$^{3}$Advanced Materials Laboratory, Sumitomo Electric Industries, Ltd. \\
1-1-1 Koyakita, Itami, Hyogo 664-0016, Japan \\

$^{4}$Graduate School of Engineering, Osaka University \\
2-1 Yamadaoka, Suita, Osaka 565-0871, Japan \\

$^{5}$Korea Institute of Ceramic Engineering and Technology \\
101 Soho-ro, Jinju-si, Gyeongsangnam-do, 52851, Korea
}

\begin{abstract}
Calculating the melting temperature, $T_{m}$, by molecular dynamics (MD) simulations is a controversial issue. Even first-principles molecular dynamics (FP-MD) simulations often give large overestimations. The accuracy of the $T_{m}$ value obtained by FP-MD simulations has been assessed by examining various solids and calculation parameters. Because the equilibria of liquids are achieved by the balance between many atom relaxation processes, adiabatic simulation---no use of thermostat---is necessary. The present simulations resolved the presence of the width $W_{m}$ for the transition, which vanishes only at the thermodynamic limit. This brings uncertainty to the obtained $T_{m}$ value by $W_{m}$.
Within this uncertainty, it has been demonstrated a good agreement in $T_{m}$ is obtained for Si and Na by a reasonable size of cells, namely, approximately 100 atoms. This implies that the contribution of surface in determining $T_{m}$ does not constitute a universal reason for use of the solid-liquid coexistence method for bulk solids, unless too large surface energy or other factors are involved. On the other hand, typically for oxide materials, the FP-MD simulations give large overestimations for $T_{m}$, which are insensitive to the calculation conditions. It is highly likely that the overestimation in $T_{m}$ is brought about by the LDA/GGA error in the electron correlation functional.
\end{abstract}

\maketitle

%%%%%%%%%%%%%%%%%%%%%%%%%%%%%%%%%%%%%%%%%%
\section{Introduction}
\label{sec:Introduction}
Numerous successes of density-functional theory (DFT) calculations in recent materials research promote the status of DFT calculations to almost an exact science. On this basis, first-principles (FP) molecular dynamics (MD) simulations, which are based on DFT, would be expected to accurately predict the melting temperature ($T_{m}$). In real studies, however, the situation is not simple. Often FP-MD simulations exhibit hysteresis, which makes it difficult to determine uniquely the melting temperature. More seriously, they often give large discrepancies in $T_{m}$, mostly overestimation \cite{Morris94,Belonoshko94,Alfe05}. The discrepancies depend on the material to be studied. In order to fix the discrepancy, new methods were proposed. Then, the result has the method dependence. This means that the study of a particular material requires the suitable method. This situation is clearly undesirable for material research. To this extent, accurate calculation of $T_{m}$ is so difficult even for DFT.

Many researchers studying melting phenomena continue to use classical MD simulations with model potentials, because of needs for prolong and large-scale simulations. The success of such simulations largely depends on the accuracy of the model potentials, which are usually tested by comparing selected properties of the material between calculation and experiment. As such, structural parameters are most commonly used. Phonon calculations using the force constant model can be looked upon as an example of construction of model potential. Often only few parameters can give a good agreement with an experimental phonon spectrum \cite{Bruesch82-1}. However, agreement in one case does not guarantee agreement with respect to other properties. As an example, Porter {\it et al.}~demonstrated that different models that produce almost identical phonon spectra predicted completely different thermal expansion values \cite{Porter97}. This inconsistency could be improved by increasing the number of parameters in a model. Even so, in the case of phenomena for which the harmonic approximation provides an inaccurate prediction, it remain unclear whether a suitable model potential can be constructed based on a reasonable number of parameters. Melting involves increased uncertainty, in that no reliable model superior to the existing Lindemann rule is available \cite{Grimvall74, Granato10}.

Reviewing previous studies concerning the melting behavior of silica-based materials illustrates the challenge associated with devising satisfactory models of potential. There have been many attempts to devise potentials for silica so as to accurately reproduce their physical properties. 
By constructing interatomic potentials of Si-O pairs in order to match the experimental values, Yamahara {\it et al.}~were able to correctly reproduce the negative thermal expansion of $\beta$-cristobalite and the $\alpha$-$\beta$ transition of cristobalite of approximately 500 K \cite{Yamahara01}. However, the same potentials produced a $T_{m}$ value for $\beta$-cristobalite about 5000 K, which is much higher than the experimental value of 1996 K. The same MD simulations also largely overestimated for the glass transition temperature, $T_{g}$, giving a value of 3800 K compared with the experimental result of 1480 K. Predicting the glass transition temperature of silica is an especially challenging task. 
Although many MD simulations have been reported \cite{Vollmayr96,Yamahara01,Kuzuu04,Takada04,Carre08,Geske16,Niu18}, the calculated values of $T_{g}$ are scattered over a wide range of temperatures from 2000 to 4000 K. Takada, {\it et al.}~report that popular potentials TTM and BKS reproduce the structures of silica well, but provide unsatisfactory $T_{g}$ \cite{Takada04}. Belonoshko expresses his surprise as ``it is puzzling why the interaction model of Matsui is successful in producing $PVT$ properties and solid-solid transformations but fail to give proper results for the pressure dependence of the melting temperature" \cite{Belonoshko94}.
Here it is helpful to note that the structural parameters that are matched with experiment when potentials are constructed are the time-averaged structures, such as the bond length. RDF is in this category. The distinction between the solid and liquid states is smeared out in those averaged quantities. The time correlation becomes more important for liquids, as discussed in this paper. We recall the often-claimed leaning that real potentials have highly non-analytic forms, which explains why the DFT approach is so valuable. 

DFT should provide, in principle, the correct answer to this problem, because the potentials are constructed from very fundamental principles of physics. However, in practice, use of DFT does not automatically guarantee the correctness of the results. 
First, even if the potential were exact, the final result will only be correct in the case that the MD simulations are properly set up. The most serious problem in this respect is imposing the periodic-boundary conditions. Melting involves a break-up of the periodicity of crystal, and thus the imposing the periodic-boundary conditions causes various spurious effects. Such effects are well studied for the defect physics: reviews are given by, for example, \cite{Castleton09,Freysoldt14}. However, different types of consideration are required for the melting phenomenon, which has not been well studied to date. 
A study by Hong and van de Walle is an example of such a few exceptions \cite{Hong13}. Based on their analysis on the size dependence, they proposed a new method to remove the hysteresis and the overheating problem in small system size. 
% Unfortunately, many factors affect the accuracy of $T_{m}$. One of them is discussed next.
% The present study is complementary to their study in that it demonstrates the case in which an accurate $T_{m}$ can be obtained without help of the coexistence method. More importantly, our inclusion of oxide materials arouses a new issue, which is described next.
% and so melting simulations are performed using supercells large enough to eliminate the effect of this artificial setup. In reality, the practical limit is a few times the size of the primitive unit cell. 

Second, the currently available DFT potentials are still approximations. The exact energy functional in DFT is not known, and will likely not be found in the near future. The overbinding problem of LDA/GGA is a well-known problem with these approximations \cite{Jones89,Parr-Yang89}. However, the influence of the overbinding on the melting temperature is not well studied \cite{note-checkDFT-liquids}. This requires calculation of the internal energy of liquid. Although calculation of energy itself is easy in DFT, it is not a trivial task for liquids. Liquids exist only at finite temperatures, and hence the calculation of $E$ without specifying $T$ has no sense. We need to determine both $E$ and $T$ as a set. Thus the problem is equivalent to calculating specific heat $C$, for which the standard theory is lacked for liquids \cite{Granato02,Trachenko11,Bolmatov12,Trachenko16,Proctor20,Baggioli21}. Determination of the relationship between $E$ and $T$ is critical. Energy dissipation (atom relaxation) processes important in determining the $E-T$ relation, and careful consideration of this effect is required for the set up of MD simulation. This is the topic that the present authors have recently established for the glass transition \cite{Shirai22-SH,Shirai22-Silica}. By utilizing the method of calculating $E$ and $T$, we are able to discuss the effect of overbinding on $T_{m}$.

As noted at the beginning, the standard method of FP-MD simulation sometimes fails to give reasonable $T_{m}$ values. In order to fix this problem, several methods were proposed, depending on what cause is envisaged as the discrepancy in $T_{m}$. These include thermodynamic integration \cite{Mei92,Sugino95,Frenkel96,Cheng19}, the solid-liquid (two-phase) coexistence method \cite{Morris94,Belonoshko94,Belonoshko01,Alfe05,Usui10,Hong13,Hong15,Geng24}, and the Z method \cite{Belonoshko06,Alfe11}.
By observing that melting is initiated from the surface of a solid, the overestimation of $T_{m}$ in MD simulations is often ascribed to the lack of surface in the standard setup of MD simulation. The two-phase coexistence method was introduced to suppress the superheating due to the lack of surface. 
Certainly, the surface has influence on determination of $T_{m}$ \cite{Cahn86}. An evident example is the reduction of $T_{m}$ in small-size systems, such as gold nanoparticles \cite{Buffat76}. However, we need to distinguish superheating of solid from supercooling of liquid. These two processes are not in the relation of reversal processes, which are discussed seriously \cite{Lu98,Jin01,Bai05,Belonoshko06,Alfe11}.
For supercooling of liquid, the mechanism of supercooling is well described by the standard theory of crystal nucleation. The surface energy $\gamma_{s}$ creates an energy barrier in the free energy $\Delta G^{\ast}$ for the crystal nucleation, which develops as $\Delta G^{\ast} = const \ \gamma_{s}^{3} (T_{m}/H_{m})^{2} /\Delta T_{m}^{2} $, $H_{m}$ is the melting enthalpy: for example, see \cite{Porter-PT-metals} (p.~192). The crystal nucleation mostly occurs at surface, which is known as inhomogeneous nucleation. In this sense, the two-phase coexistence method is useful to help the creation of surface nucleation in practically realizable times of simulation.
In contrast, for superheating of solid, it is not clear what the substance of energy barrier is. There has been, so far, no direct evidence for the ``liquid nucleation". 
By using the two-phase coexistence method, Belonoshko demonstrated a reduction in the calculated $T_{m}$ by more than 1000 K for MgO \cite{Belonoshko94}. It is suspicious that the surface barrier has such a large effect on $T_{m}$ for bulk solids; the supercooling can be normally achieved with the reduction $\Delta T_{m}$ by less than 10 \% of $T_{m}$.

Generally speaking, the bonding environments of the surface atoms are different from that of bulk atoms. Accordingly, it is reasonable to observe the surface effect for small-size systems, such as nanoparticles. The surface effect can be scaled with the inverse of the system size $L$, which presents the ratio of the number of surface atoms ($N_{S}$) to the number of bulk atoms ($N_{V}$). With a suitably chosen $L$, the reduction of the melting temperature, $\Delta T_{m}$, is given by
\begin{equation}
\frac{\Delta T_{m}}{T_{m}} \propto \frac{1}{L} \propto \frac{N_{S}}{N_{V}}.
\label{eq:scale-surface}
\end{equation}
See Eq.~(14) in \cite{Buffat76}. In view of this scaling, the current size of the two-phase coexistence method is of the same order of the magnitude that nanoparticles have. This raises the doubt that the reduction in $T_{m}$ by the coexistence method could be due to the size effect. If this is the case, the result of the coexistence method turns to be underestimation for the bulk. We need to discuss the overestimation in calculations separately from the superheating of experiment. We should not overlook the possibility that the overestimation in $T_{m}$ is due to calculation errors caused by, for example, use of the small cell size or use of inaccurate potentials. 
In any respect, neither the mechanism of superheating nor the assessment of the coexistence method is the purpose of the present study. Here, we are interested only in the accuracy of $T_{m}$ in the standard setup of FP-MD simulations.

In this paper, the accuracy of $T_{m}$ values obtained by FP-MD simulations is assessed by the method employed in previous studies \cite{Shirai22-SH,Shirai22-Silica}. In Sec.~\ref{sec:Method}, the characteristics of the method is explained. It is emphasized that the energy-dissipation processes are essential in determining $T_{m}$.
In Sec.~\ref{sec:Tm}, we report the $T_{m}$ values obtained by FP-MD simulations for various materials. The result shows that the overestimation in $T_{m}$ often occurs.
The reason for the overestimation is not unique and depends on material. Further analysis is made for selected materials in the remaining part of Sec.~\ref{sec:Results}. These examples are chosen from semiconductors, metals, and oxides. This selection is quite arbitrary, and there is no reason that the accuracy is in accordance with the material classification. 
In Sec.~\ref{sec:Conclusion}, the present results are summarized.

%%%%%%%%%%%%%%%%%%%%%%%%%%%%%%%%%%%%%%%%%%
\section{Methods}
\label{sec:Method}
\subsection{Simulation of melting}
\label{sec:Sim-melt}
This work employed both dynamical and thermodynamical methods to determine $T_{m}$ on the basis of MD simulations. The most commonly used method may be the calculation of diffusion constant $D$. $T_{m}$ is determined as the temperature at which a finite value $D$ appears as a solid is heated. The calculation of $D$ is a quite standard method \cite{Frenkel96}, and thus no explanation is needed here. The other method is the calculation of the internal energy $U$, because $U(T)$ exhibits an abrupt change at $T_{m}$. This, in turn, provides the latent heat associated with the melting enthalpy $H_{m}$. Because of the importance of energetics, we focus more on this $U$ calculation.
Recently, the authors developed a method of calculating the specific heat associated with the glass transition \cite{Shirai22-SH,Shirai22-Silica}. This method is advantageous in that it provides the specific heat and other thermodynamic functions for both solids and liquids with an equal level of accuracy. This method is described below in order to emphasize the full thermodynamic treatment of the liquid state.

\subsubsection{Energy versus temperature}
\begin{figure}[ht!]
\centering
    \includegraphics[width=120mm, bb=0 0 650 450]{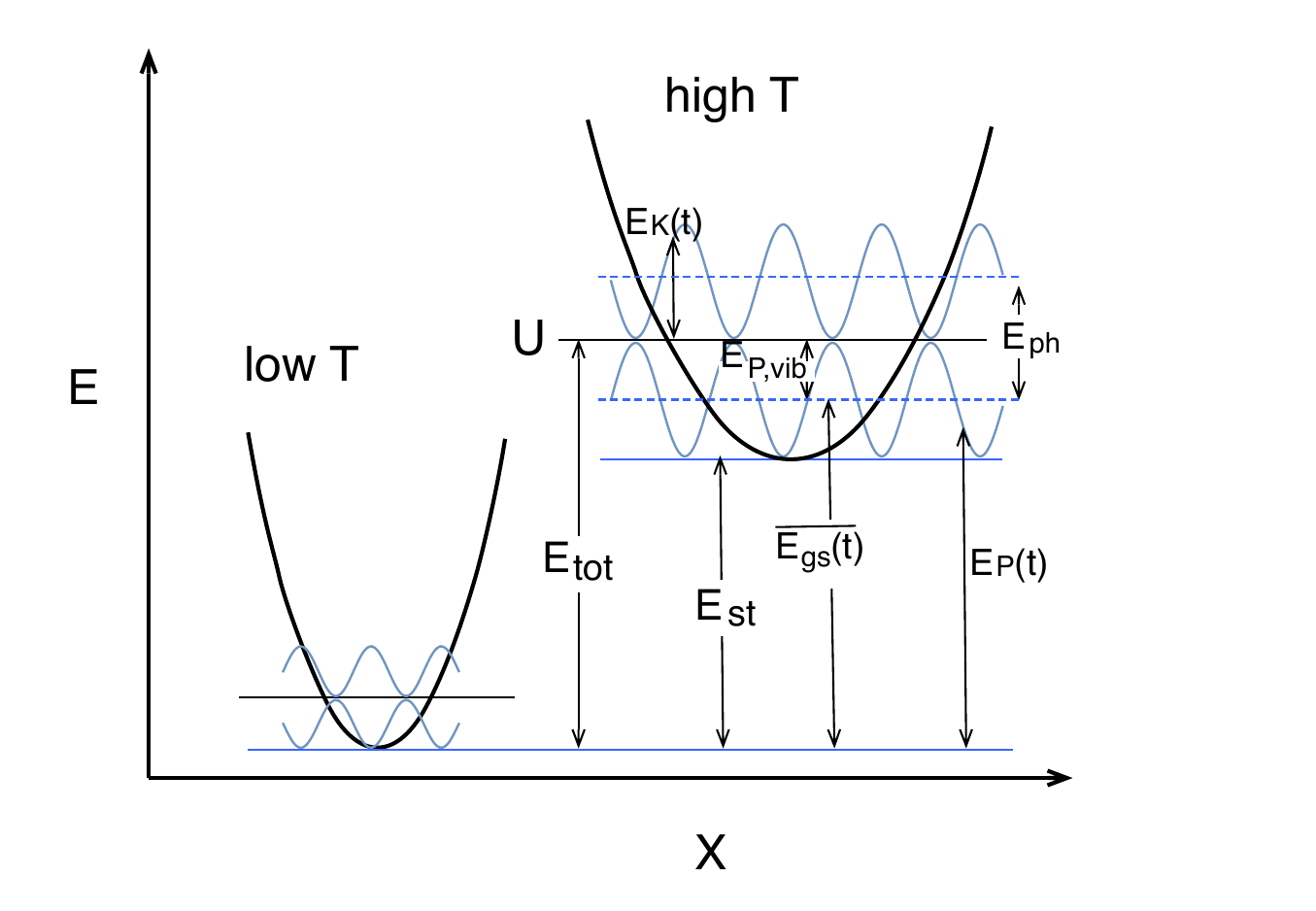} 
\caption{
The relationships between various energies in adiabatic MD simulations. As $T$ increases, the bottom of the adiabatic potential is raised by $E_{\rm st}$.
}
\label{fig:E-Xcurve} 
\end{figure}

Let us consider an MD simulation for a system composed of $N$ atoms. The $j$th atom having mass $M_{j}$ is at position ${\bf R}_{j}(t)$ with velocity ${\bf v}_{j}(t)$ at time $t$. On the Born-Oppenheimer approximation, the total energy, $E_{\rm tot}$, of a system is given by the sum of the kinetic energy, $E_{\rm K}$, and the potential energy, $E_{\rm P}$, as
\begin{equation}
E_{\rm tot}(t) \equiv  E_{\rm P} + E_{\rm K} = 
    E_{\rm gs}( \{ {\bf R}_{j}(t) \}) + \frac{1}{2} \sum_{j} M_{j} v_{j}(t)^{2},
\label{eq:total-energy}
\end{equation}
where $E_{\rm gs}( \{ {\bf R}_{j}(t) \})$ is the ground state energy of the electrons, including ion-ion potentials, for the instantaneous positions $\{ {\bf R}_{j}(t) \}$.
The internal energy in thermodynamics is defined at equilibrium and is given by the time average of the microscopic total energy $E_{\rm tot}(t)$ of the system,
\begin{equation}
U = \overline{E_{\rm tot}(t)} = \overline{E_{\rm gs}( \{ {\bf R}_{j}(t) \})}
    + \frac{1}{2} \sum_{j} M_{j} \overline{ v_{j}(t)^{2} }.
\label{eq:internal-energy}
\end{equation}
For solids, the constituent atoms fluctuate around their equilibrium positions. The instantaneous position of the $j$th atom can be expressed by the sum of the equilibrium position, $\bar{\bf R}_{j}$, and a small displacement from this position, $\bar{\bf u}_{j}$, as ${\bf R}_{j}(t) = \bar{\bf R}_{j} + {\bf u}_{j}(t)$. The ground-state energy, $E_{\rm gs}( \{ {\bf R}_{j}(t) \} )$, can be expanded in the Taylor series in terms of this displacement. The part of time dependent term together with $E_{\rm K}$ constitutes the phonon energy and its time average $E_{\rm ph}$ is given by
\begin{equation}
E_{\rm ph} = \frac{1}{2} \sum_{i,j} \overline{ {\bf u}_{i}(t)\cdot {\bf D}_{ij} \cdot {\bf u}_{j}(t)}
    + \frac{1}{2} \sum_{j} M_{j} \overline{ v_{j}(t)^{2} },
\label{eq:phonon-energy}
\end{equation}
where ${\bf D}_{ij}$ is the dynamic matrix between the $i$th and $j$th atoms \cite{BornHuang}. The first term of the right-hand side of Eq.~(\ref{eq:phonon-energy}) is referred to $E_{\rm P, vib}$ herein.
The remaining part, $E_{\rm st}( \{ \bar{\bf R}_{j} \} )$, in the expansion $\overline{E_{\rm gs}( \{ {\bf R}_{j}(t) \})}$ is constant with respect to time and is referred to as the {\em structural} energy. Thus, the internal energy $U$ consists of the following terms,
\begin{equation}
U \equiv U(T,V,\{ \bar{\bf R}_{j} \} ) = E_{\rm st}( \{ \bar{\bf R}_{j} \} )  +  E_{\rm ph}(T) + E_{\rm te}(V).
\label{eq:internal-energy-3}
\end{equation}
In Eq.~(\ref{eq:internal-energy-3}), three translation and three rotational displacements of the entire system have been removed from the term $E_{\rm st}( \{ \bar{\bf R}_{j} \} )$. The contribution of these displacements is extracted to form the last term $E_{\rm te}(V)$, which is referred to as the thermal expansion energy. Therefore, the structural energy $E_{\rm st}$ means the energy change by the internal coordinates with fixed lattice parameters. 
Figure \ref{fig:E-Xcurve} summarizes the relationships between $U$ and these components.

By diagonalizing the dynamic matrix, $E_{\rm ph}$ can be calculated as
\begin{equation}
E_{\rm ph} = \int \hbar \omega \left( \bar{n}(\omega) + \frac{1}{2} \right) g(\omega) d\omega,
\label{eq:phonon-energy-integral}
\end{equation}
where $\hbar$ is the Planck constant, $\bar{n}(\omega)$ is the Bose occupation number, and $g(\omega)$ is the phonon DOS. Because atom motions are treated classically in conventional MD simulations, the kinetic energy of atoms in these simulations does not obey the Bose-Einstein statistics but instead obeys the classical equipartition law of energy,
\begin{equation}
\overline{E_{\rm P,vib}} = \overline{E_{\rm K}} = \frac{3}{2} N k_{\rm B}T,
\label{eq:equi-partition}
\end{equation}
where $k_{\rm B}$ is Boltzmann's constant. In order to fix this problem, first this classical term in $\overline{E_{\rm gs}(t)}$ is removed from the total energy, giving the structural energy $E_{\rm st}$,
\begin{equation}
E_{\rm st} = \overline{E_{\rm gs}(t)}-\overline{E_{\rm K}} = \overline{E_{\rm gs}(t)} - \frac{1}{2} E_{\rm ph}.
\label{eq:Est}
\end{equation}
Then, the phonon energy $E_{\rm ph}$ is evaluated through Eq.~(\ref{eq:phonon-energy-integral}). The phonon DOS, $g(\omega)$, is obtained from the power spectrum of atom velocities in an MD run \cite{Payne92}, whereas the temperature in the Bose occupation number is determined by the kinetic energy of atoms via Eq.~(\ref{eq:equi-partition}). Finally, $E_{\rm tot}$ is recalculated by adding this phonon energy $E_{\rm ph}$ to $E_{\rm st}$, recovering the Bose-Einstein statistics.
In practice, this modification of $E_{\rm tot}$ is important only at very low temperatures.

For liquids, the atoms do not have unique equilibrium positions and hence the expansion (\ref{eq:phonon-energy}) does not make sense. Despite this, we can operationally define a ``phonon" energy for a liquid in a manner similar to that leads to Eq.~(\ref{eq:phonon-energy-integral}). This is because the frequency spectrum of atom velocities, $f_{v}(\omega)$, can be obtained from MD simulations. In fact, deriving the phonon dispersion in this manner is a practice in the inelastic neutron scattering analysis \cite{Egelstaff-2ed,Copey74,Smith17}. Observation of even a Brillouin-zone-like structure is surprising \cite{Suck05}. The frequency spectrum of atom velocities $f_{v}(\omega)$, however, does not distinguish scattering from vibrational motions. The idea of $f_{v}(\omega)$ can be applied even for an ideal gas, in that collisions create oscillatory behavior. This phonon-like scenario is often used in modeling liquids, which often gives a good description of the specific heat $C_{p}$ \cite{Wallace97b,Wallace98, Bolmatov12,Trachenko16,Baggioli21}. We note, however, that the specific heats of many metal liquids are, from beginning, close to the classical limit of the equipartition law of energy, $C_{v} = 3 k_{\rm B}$ \cite{Wallace02}, which is insensitive to the mode of atom motions. Throughout this paper, the specific heat per atom is understood by using the unit $k_{\rm B}$. Upon closer observation, it will be seen that the phonon-like model described above is only an approximation, as discussed in Sec.~\ref{sec:Silicon}. A good example is water, for which $C_{p} \cong 3 k_{\rm B}$ even though H-related vibrations cannot be thermally activated at the $T$ range of relevance. Liquid Se has a large $C_{p}$ more than 4 $k_{\rm B}$ \cite{Shu80}.
In many metallic liquids, $C_{v}$ near $T_{m}$ falls in a range 3.0 to 3.4 $k_{\rm B}$, which already exceeds the classical limit (Ref.~\cite{Wallace02}, p.~242). Some authors ascribe this excess specific heat to the anharmonic effect of phonons \cite{Wallace98, Bolmatov12,Trachenko16, Proctor20,Baggioli21}. But, this excess specific heat increases as $T$ is lowered to $T_{m}$ from the high temperature side. The anharmonic effect of phonons is expected to be reduced upon lowering $T$ \cite{Leibfried61}. Moreover, recent experiments on supercooled liquids show that the increasing behavior of the excess specific heat as $T$ decreases is continuously extended in the range $T<T_{m}$, namely, $dC_{p}/dT \approx $ const with a negative constant value: \cite{Perepezko84} for metallic liquids, \cite{Rhim00a, Li04} for melton Ge. Therefore, it is unrealistic to ascribe the increasing behavior of the excess specific heat with decreasing $T$ to the anharmonic effect of phonons. 

% ostensible
In spite of facile success of phonon theory for liquids, the theory eventually fails to accurately describe liquid states \cite{note-phonon-liquids}. Around the transition temperature of phase transition, the contribution of $E_{\rm st}$ plays the predominant role in the $T$ dependence on $U$, as shown by previous MD simulations \cite{Shirai22-SH,Shirai22-Silica}. Recent studies on highly viscous glass-forming materials, show that the configuration contribution---this corresponds to $E_{\rm st}$ in this paper---dominates the excess entropy in the supercooled liquid state \cite{Smith17,Alvarez-Donado20,Han20}. The contribution from $E_{\rm st}$ must therefore be taken into account. In the present method, the total energy $E_{\rm tot}$ includes both the components $E_{\rm st}$ and $E_{\rm ph}$. In the sprit of DFT, the total energy is the most reliable quantity.

The difficulty in calculating the thermodynamic functions of liquids lies essentially in that there is no eigenstate for liquids. There is no {\em microscopic} steady state. In standard applications of statistical mechanics, the presence of eigenstates $\{ i \}$ is presumed, when the partition function, $Z=\sum_{i} e^{-\varepsilon_{i}/k_{\rm B}T}$, is evaluated. In the phonon model for liquids, phonon-like excitations are envisaged. But this phonon is soon destroyed after creation. The energy conservation requires that the destruction of a phonon is compensated for by creation of new phonons \cite{Leibfried61}. The liquid state can therefore be considered as a {\em macroscopic} steady state achieved by the dynamic balance between incessant creation and destruction of phonon-like excitations. 
In an adiabatic simulation in a sense that no heat bath is involved, the total energy of the system is conserved and the temperature is adjusted to achieve this dynamic balance. The importance of the energy dissipation in determining $T$ is demonstrated by the frequency dependence of the specific heat in the glass transition \cite{Birge85,Nielsen96,Hentschel08}. 
By construction, this energy dissipation process is fully described by MD simulations because of the rigorous equations of atom motions, and hence the correct $U-T$ relationship can be automatically obtained by MD simulations. However, if the viscose term coupled to the thermostat is introduced into the MD simulation, the energy dissipation processes are modified by this artificial coupling between the system and the thermostat, destroying the intrinsic relationship between $U$ and $T$. The melting temperature is determined by subtle balance among many energy dissipation processes. Therefore adiabatic simulations are essential to obtain the correct $U-T$ relationship.\cite{note-Relax} In the glass transition, the effect of atom relaxation is well known to yield hysteresis and complicates the analysis particularly when temperature is scanned in caloric methods. To remove the effect of relaxation, the adiabatic calorimetry is used \cite{Davies53,Davies53a}. To emphasize this feature, we use the term adiabatic MD simulation in place of the widely-used term $NVE$ simulation. How adiabatic methods separate the hysteresis from the time dependence properties are discussed in \cite{Shirai24-hysteresis}.

\subsubsection{Construction of MD simulations}
In this study, the internal energy $U$ must be evaluated over a wide range of temperature to cover from the solid to liquid states. To ensure that equilibrium was established at each $U$, an adiabatic MD run was performed for a fixed $E$ and repeated by changing the total energy. In adiabatic MD simulations, the control parameter is the initial temperature $T_{\rm in}$, at which the initial velocities of atoms are given through Eq.~(\ref{eq:equi-partition}). In each single MD run, the simulation was continued until equilibrium was established. It is noted that there are two kinds of equilibrium conditions \cite{Shirai20-GlassState}. The one is the equilibrium condition with respect to temperature (thermal equilibrium). The equilibrium temperature is determined by
\begin{equation}
\frac{3}{2} k_{\rm B}T = \frac{1}{2} \langle M_{j} \overline{ v_{j}(t)^{2} } \rangle_{j}.
\label{eq:equil-temp}
\end{equation}
Here, the brackets denote the particle average, whereas the bar indicates the time average.
Thermal equilibrium is reached when the time average of $E_{\rm K}(t)$ converges. In the following, $T$ in all figures indicates this equilibrium temperature. The time required to reach this condition is referred to as the thermal relaxation time, $\tau_{T}$. For most of solids, $\tau_{T}$ is quite short, on the order of 0.1 ps. For liquids, $\tau_{T}$ becomes longer but is still less than 1 ps. Thus, in most cases the relaxation to equilibrium is controlled by the other condition, namely, the mechanical (structural) relaxation. The time required to this relaxation is referred to as the structural (mechanical) relaxation time, $\tau_{M}$.
The time evolution of the particle-averaged displacements, 
\begin{equation}
\langle \overline{\delta R_{j}(t)^{2} } \rangle_{j} = 
\frac{1}{N t_{\rm sm}} \sum_{j}^{N} \int \left( R_{j}(t-t_{0}) - R_{j}(t_{0}) \right)^{2} dt_{0},
\label{eq:average-displace}
\end{equation}
is used to determine whether structural equilibrium is reached, where $t_{\rm sm}$ is the simulation time. For solids, it is considered an equilibrium state when $\langle \overline{\delta R_{j}(t)^{2} } \rangle_{j}$ is constant with respect to $t$. In contrast, for liquids, it is considered an equilibrium state when $\langle \overline{\delta R_{j}(t)^{2} } \rangle_{j}$ shows a linear dependence on $t$ over the entire simulation time $t_{\rm sm}$. For the latter case, the slope, $D = (1/6) \lim_{t \rightarrow \infty} \langle \overline{\delta R_{j}(t)^{2} } \rangle_{j} /t$, gives the diffusion coefficient $D$. For the liquids treated in this study, $\tau_{M}$ was on the order of a few ps. Examples of the time development of $E_{\rm K}$ and $\langle \overline{\delta R_{j}(t)^{2} } \rangle_{j} $ are provided in the Supplemental Material to Ref.~\cite{Shirai22-Silica}.

\subsection{Calculation conditions}
\label{sec:condition}
While several codes were used for the present FP-MD simulations, the majority of calculations were performed by Phase/0 \cite{PHASE}. Some were performed by a code written in-house (Osaka2k) or the VASP code \cite{VASP}. All codes are pseudopotential methods using planewave expansion.
The MD simulations employed time steps ranging from 0.72 to 1.2 fs and the total simulation time, $t_{\rm sm}$, varied from 2 to 10 ps, depending on the relaxation time. The volume of the crystal, $V$, was fixed at the experimental value, although normally the $V$ of the corresponding liquid phase is slightly larger than that of the crystal. The effect of volume change on $T_{m}$ is examined in subsections \ref{sec:Na} and \ref{sec:silica}.

%%%%%%%%%%%%%%%%%%%%%%%%%%%%%%%%%%%%%%%%%%
\section{Results and discussion}
\label{sec:Results}

\subsection{Melting temperature}
\label{sec:Tm}

Melting simulations for various materials were performed by FP-MD simulation by scanning a wide range temperature covering both the solid and liquid states. Table \ref{tab:sum-Tm} summarizes the calculated $T_{m}$ together with experimental values for comparison. The potential type is indicated by NC (norm-conserving), US (ultrasoft), PAW (projector augmented waves) \cite{Martin04}, along with the type of electron correlation energy functional (LDA or GGA). 
In MD simulations, melting occurs over a finite range of temperature, as shown later. Here this temperature range is referred to as the {\it transition region}. The calculated value of $T_{m}$ is then defined as the mid point of the transition region. The width, $W_{m}$, of the transition region is expected to approach zero as the size of supercell is increased. 
In the table, the deviation of the calculated value from the experimental value is indicated by the ratio, $r$, of the former to the latter. In all the cases studied, $r>1$ was obtained, meaning that the calculated values were always overestimated. It is our surprise to obtain large values of $r$ more than 2, despite the initial confidence in the DFT calculations. Large overestimations of $r>2$ are found when the material contains oxygen atoms. The involvement of oxygen for the large overestimation may be a reason why the coexistence method is widely used in the field of geophysics, because most materials important for that field are oxides \cite{Belonoshko94,Alfe05,Usui10,Paola16,Hernandez22,Geng24}.
As$_{2}$Se$_{3}$ also has a large $r$. A recent study on c-BAs reports a large improvement in $T_{m}$ using the coexistence method \cite{Cheng24}. The large overestimations of As$_{2}$Se$_{3}$ and of c-BAs could have the common origin.
Only for Si, the calculated value is in a good agreement with experiment. Note that, in all the calculations in Table \ref{tab:sum-Tm}, the used volume of the unit cell is that of the crystal, which was obtained by structural optimization at $T=0$. Later, the influence of volume is investigated.

In the table, the melting enthalpy $H_{m}$ is also listed in units of eV/atom. $H_{m}$ is obtained from the abrupt change in $E_{\rm st}$. In experiment, the change $H_{m}$ appears at a discrete temperature $T_{m}$, whereas melting occurred over a finite width $W_{m}$ in the present MD simulations. Hence, the accuracy of $H_{m}$ depends on the sharpness of the transition region. Small cell sizes give less reliable $H_{m}$ values.
The accuracy of $H_{m}$ highly depends on the accuracy of $T_{m}$. These two quantities are related each other by
\begin{equation}
H_{m} = T_{m} \Delta S_{m},
\label{eq:Hm-Tm}
\end{equation}
where $\Delta S_{m}$ is the entropy of melting. This relationship says that an error of 0.1 eV in $H_{m}$ causes an error of 1000 K in $T_{m}$, provided $\Delta S_{m}/k_{\rm B} = 1$. Because $\Delta S_{m}$ in many materials is of this order of magnitude, the accuracy of $H_{m}$ directly controls the accuracy of $T_{m}$. Again the calculation for Si shows a good agreement in $H_{m}$ with experiment. This confirms the internal consistency between calculated $T_{m}$ and $H_{m}$ for Si. The phonon energy, $E_{\rm ph}$, and thermal expansion energy, $E_{\rm te}$, can contribute $H_{m}$. However, the previous studies \cite{Shirai22-SH,Shirai22-Silica} showed that these two contributions can generally be ignored, unless there is a large difference in thermal expansion between the solid and liquid phases. Because the melting temperature of most materials is high, the specific heat of the phonon contribution near $T_{m}$ becomes already the classical limit of the Dulong-Petit law, $C_{v} = 3 k_{\rm B}$, and thus there is no reason that an abrupt jump in $E_{\rm ph}$ occurs at $T_{m}$.

\begin{table}
\begin{tabular}{l  | rrc | rr | r | l rr }
  \hline \hline
Material  & \multicolumn{3}{c|}{ $T_{m}$ (K) } & 
\multicolumn{2}{c|}{$H_{m}$ (eV/atom)} & Size & \multicolumn{3}{c}{Potentials} \\
   \cline{2-4} \cline{5-6} 
   & exp. & calc. & $r$ &  \multicolumn{1}{|c}{exp.} & calc.  & (atoms) & \multicolumn{1}{l}{type} & $E_{\rm cut}$ & code \\ 
   \hline
Si & 1683 & 1750 & 1.04 & 0.52 & 0.41 & 512 & NC+GGA & 204 & Phase \\
Na & 371 & 460 & 1.24 & 0.026 & 0.015 & 128 & NC+GGA & 245 & Phase \\
$\alpha$-quartz  & 1700 & 4500 & 2.64 & 0.033 & 0.13 & 72 & US+GGA & 408 & Phase \\
 \quad (SiO$_{2}$) & & & & & &  & & \\
Glycerol  & 291 & 635 & 2.18 & 0.013 & 0.010 & 56 & US+GGA & 408 & Phase \\
 \quad (C$_{3}$H$_{8}$O$_{3}$) & & & & & &  & & \\
B$_{2}$O$_{3}$ & 723 & 3000 & 4.15 & 0.051 & 0.18 & 60 & PAW+GGA & 340 & Phase \\
H$_{2}$O & 273 & 720 & 2.63 & 0.021 & 0.03 & 96 & NC+LDA & 816 & Osaka2k \\
As$_{2}$Se$_{3}$ & 645 & 1080 & 1.67 & 0.037 & 0.1 & 160 & US+GGA & 408 & Phase \\
   \hline
\end{tabular}
\caption{Comparison of calculated $T_{m}$ to the experimental value for various materials. $r$ indicates the ratio of the calculated $T_{m}$ to the experimental value. The melting enthalpy $H_{m}$ is also listed. Size indicates the size of the supercell used in MD simulations. The last block indicates the used potentials. $E_{\rm cut}$ for planewave expansion is given in eV units. $\Gamma$-point $k$ sampling is used in all the listed data. Experimental data are from \cite{CRC92} with the exception of the data for quartz \cite{Richet82}. }
\label{tab:sum-Tm}
\end{table}

Several causes for the error in the calculated $T_{m}$ are investigated. One important cause is the use of periodic boundary conditions. An effect of the periodic boundary conditions is elimination of the surface effects. Because melting first occurs at the surface of a material, it can be argued that elimination of the surface makes melting less likely, thus increasing $T_{m}$. This effect could contribute to the overestimation of $T_{m}$. However, based on the good agreement between the calculated and experimental $T_{m}$ values for Si, this cannot be the primary reason for the large overestimation of $T_{m}$ unless a very large surface energy is involved. The introduction of defects could have a similar effect on melting. However, in our experience for quartz \cite{Shirai22-Silica}, the introduction of a defect in a supercell did not affect $T_{m}$ within the numerical accuracy.

Another effect of the periodic boundary conditions is the creation of spurious reflections of atom movements, which yields an artificial barrier to such movements. For a given size of supercell, $L$, all long-wavelength phonons (those with wavelength $\lambda > L$) are eliminated. In a previous paper \cite{Shirai22-Silica}, by using a crude model, it is shown that this artificial barrier, $V_{0}$, is reduced only in a scale $1/L^{2}$, while the computational demands increase as $L^{3}$. The value of $V_{0}$ can be of the order of a few tens of eV with a cell size of $2 \times 2 \times 2$ for $\alpha$-quartz. However, to test this scaling, real calculations are needed by increasing the cell size. In practice, extending the size of supercells beyond those listed in Table \ref{tab:sum-Tm} is very expensive. For a 512-atoms simulation, one month computation was required for a single temperature in our cluster machines (typically in an 8-core parallel configuration). Hence, the effect of supercell size can only be examined by reducing the supercell size.
In the following, the effect of supercell size is examined by taking one example from each of the categories of semiconductors, metals, and oxide compounds. All the detailed data listed in Table \ref{tab:sum-Tm} are seen in the Supplemental Materials.

% -------------------------------------------------
\subsection{Silicon}
\label{sec:Silicon}
\begin{figure}[htbp]
  \centering
     \includegraphics[width=120 mm, bb=0 0 424 480]{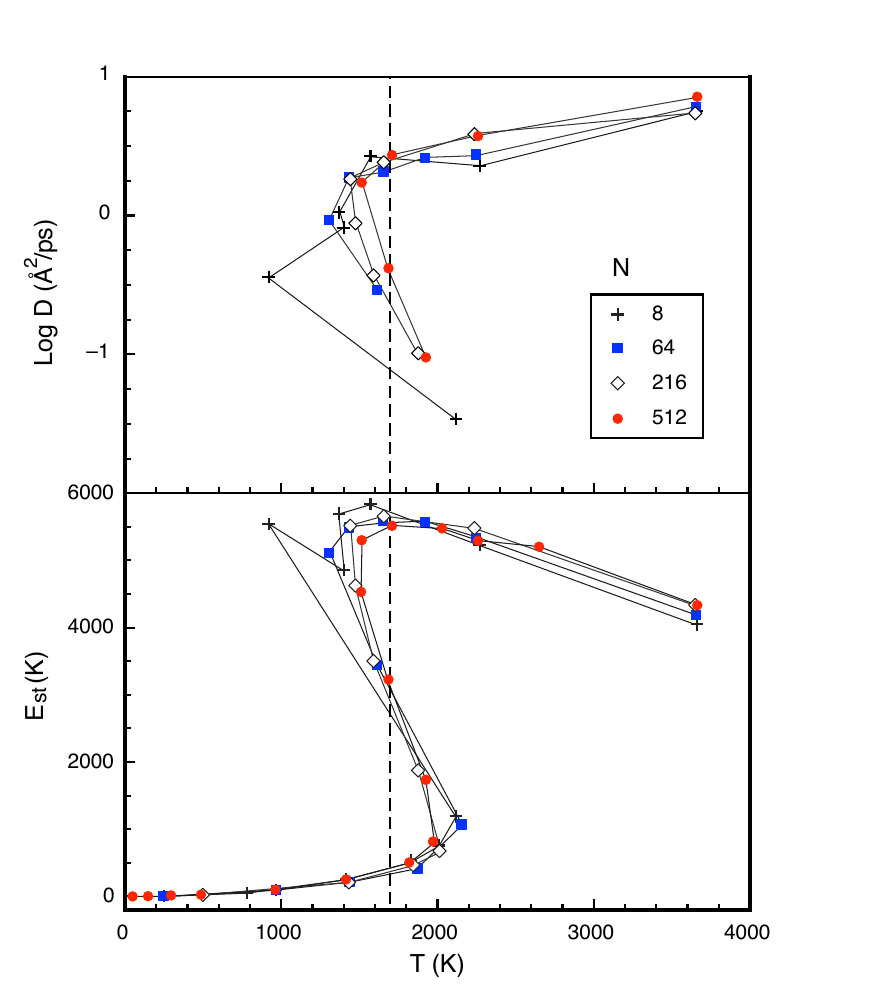} 
  \caption{The diffusion coefficient, $D$, and structural energy, $E_{\rm st}$, of Si calculated by NC+GGA. The cell sizes ($k$ meshes) are $N=8$ ($4^{3}$), 64 ($2^{3}$), 216 (1), and 512 (1). $E_{\rm cut}=204$ eV. The experimental value for $T_{m}$ is indicated by the dashed line. For each $N$, the energy origin is set to the ground state energy of that cell. 
  } \label{fig:Est-lnD-si}
\end{figure}

The first example is Si. The cell sizes of $N=$8, 64, 216, and 512 atoms were examined. In Fig.~\ref{fig:Est-lnD-si}, the structural energy, $E_{\rm st}$, and diffusion constant, $D$, are plotted as functions of $T$. Note that $E_{\rm st}$ is normalized by the number of atoms in the cell and is in units of K. This is convenient because the specific heat is immediately obtained in units of $k_{\rm B}$, making it easier to compare the obtained value to the classical limit $3 k_{\rm B}$. This convention is used throughout this paper. In this figure, the data points are connected by lines to show the sequence of changing $T_{\rm in}$. 
As shown in Fig.~\ref{fig:Est-lnD-si}, $E_{\rm st}$ is virtually constant against $T$ up to $T=1000$ K, indicating that the crystal potential is well described by the harmonic approximation. Finite values of $D$ appear at approximately $T=2000$ K, confirming the onset of melting. In accordance with the variation in $D$, $E_{\rm st}$ exhibits an abrupt increase there. 

The figure shows that the $E_{\rm st}-T$ curve has the width, $W_{m}$, for the melting temperature. The $E_{\rm st}-T$ curve is expected to behave as a step function at $T_{m}$ and the magnitude of step must correspond to the latent heat. This width is the artifact due to use of finite-size cells, as clearly shown by its $N$ dependence. There is an unavoidable uncertainty in temperature, $\Delta T$, for finite-$N$ systems.
\begin{equation}
\frac{\Delta T}{T} = \frac{\Delta E_{\rm K}}{ \overline{E_{\rm K}} } = \frac{1}{\sqrt{N}}.
\label{eq:fluctuaion-T}
\end{equation}
(Sometimes, it is referred to as the temperature fluctuation, but it may be better to term the kinetic-energy fluctuation \cite{Kittel88,Mandelbrot89}.) The heat exchange to the heat bath in MD simulations increases the kinetic-energy fluctuation, and this is another reason why the present method avoids the heat bath. 
This artificial width for melting is common in all materials, aside from the magnitude $W_{m}$.
Furthermore, we notice in Fig.~\ref{fig:Est-lnD-si} that the transition curves, both $E_{\rm st}$ and $D$, exhibit sigmoidal behaviors, that is, negative slopes. The negative slope in $E_{\rm st}$ against $T$ means a negative specific heat, which is thermodynamically unstable \cite{Lynden-Bell99,Landsberg87,Michaelian07}. The negative slope in $D$ against $T$ also means dynamical instability. This behavior was pointed out previously as the effect of finite-size cells \cite{Shirai22-SH,Shirai22-Silica}.
When the crystal melts, the energy distribution between the vibrational motions (in the solid portion) and translational motions (in the liquid portion) becomes unbalanced due to this energy fluctuation. At temperatures slightly below $T_{m}$, those atoms with sufficiently high kinetic energy (higher than $k_{\rm B}T_{m}$) begin to convert their vibrational motions to diffusing motions, decreasing the average kinetic energy $\langle E_{K} \rangle$ and increasing diffusion. In contrast, when this low-energy part of the distribution is overpopulated, the system temperature becomes lowered. This makes these low-energy diffusing atoms being trapped in the crystal potential, leading to a release of the energy in a form of emitting heat with a decrease in $D$. In this manner, an oscillatory behavior appears around $T_{m}$. In experiment, this sigmoidal behavior has been observed in clusters containing small numbers of atoms \cite{Schmidt21}.
The transition region is narrowed as $N$ increases. On the thermodynamic limit, $N \rightarrow \infty$, $W_{m} \rightarrow 0$ is reasonably expected. However, it is noticeable that $W_{m}$ of 300 K remains even for the cell size of 512 atoms. The convergence of the energy fluctuation is slow.
% $\langle E_{K} \rangle$,

The presence of this unstable region has some implications to the interpretation of MD simulations. When the temperature in a MD run is controlled by some means, such as thermostat, the system would happen to be trapped at either of the two extrema in the sigmoidal curve. If the system is relaxed to the maximum $T$ in the sigmoid, the obtained $T_{m}$ would become an overestimation. If the system is relaxed to the minimum $T$, then it would give an underestimation. These two extrema are spurious equilibria due to use of a finite $N$. The risk of this unstable region would be increased by using thermostats, because introduction of a thermostat usually increases the width of the energy fluctuation, even though the average temperature is well controlled.
There is possibility that spurious equilibria happen in the Z method \cite{Belonoshko06,Alfe11}. Oscillatory behaviors between solid and liquid states are reported in \cite{Alfe11}.

For Si, even though the convergence for the width $W_{m}$ is very slow, the calculated $T_{m}$, which is obtained by the mid point in the transition region, already agreed with the experimental value, when a cell size of $N=64$. Even a small cell of $N=8$ is not bad, aside from the $W_{m}$. From the viewpoint of construction of FP pseudopotential, Si is the easiest element \cite{BHS82,TM91}. The potential is not deep and there are no semi-core levels. This may be the reason why the good agreement is obtained for $T_{m}$.
In the literature, calculations of the melting temperature of silicon are reported by using thermodynamic integration \cite{Sugino95,Alfe03}. Although the agreement with the experiment is almost satisfactory, they gave an underestimation about 20 \%. This underestimation could be accounted by the above-mentioned trapping at the spurious equilibrium with the lowest minimum. Because the thermodynamic integration method, many artificial steps of calculation, such as preparation of the reference state and adiabatic increase in the interaction parameter, are involved, it is difficult to analyze at which step the error occurs. 

For the liquid state, a decrease in $E_{\rm st}$ against $T$ can be seen in Fig.~\ref{fig:Est-lnD-si}. This negative $T$ dependence is unphysical. In contrast, the total energy, $E_{\rm tot}$, is an increasing function of $T$, as should be. A full description of the $T$ dependence of $E_{\rm tot}$ will be given elsewhere \cite{Shirai-EntropyLiquid}. Here, we only mention that this decreasing behavior of $E_{\rm st}$ is a consequence of applying Bose-Einstein statistics to the ``phonons" of the liquid state. The decrease in $E_{\rm st}$ results in by subtracting overly large $E_{\rm ph}$ value from $\overline{E_{\rm gs}(t)}$ in Eq.~(\ref{eq:Est}).
This temperature range ($T>T_{m}$) is already the classical regime, in which the Dulong-Petit law is expected to hold, if the phonon model is valid. All ``phonon" modes are fully activated, which means that the thermal energy $(1/2) k_{\rm B}T$ is assigned to each mode $q$ of phonons according to Eq.~(\ref{eq:phonon-energy-integral}), for which the independent particle description for phonons is assumed. In experiment, the $C_{v}$ of liquids decreases from approximately $3 k_{\rm B}$ near $T_{m}$ to approximately $2 k_{\rm B}$ near the boiling temperature $T_{b}$ (Ref.~\cite{Wallace02}, p.~244). From the phonon theory, it can be interpreted that the degree of phonon freedom, $f$, per atom is effectively reduced from $f=6$ at $T_{m}$ to $4$ at $T_{b}$. Therefore, applying Eq.~(\ref{eq:phonon-energy-integral}) to the liquid state leads to an overestimation of the phonon energy, resulting in the decrease in $E_{\rm st}$. The independent particle description is not valid for the ``phonons" of liquids. 
This invalidity of the independent particle description for liquids can be understood by considering the extreme case of an ideal gas, where $C_{v}$ has the limiting value $f=3$, which is usually interpreted as the freedom of purely translational motions. When the frequency spectrum, $f_{v}(\omega)$, of atom velocity is calculated for an ideal gas, we have a ``phonon" spectrum, $g(\omega)$. Even though there is no restoring forces, atom collisions create the oscillatory behavior, yielding $f_{v}(\omega)$. Obviously, in this case, the phonon energy, Eq.~(\ref{eq:phonon-energy-integral}) should not contribute to the $C_{v}$ of the gas.
On this basis, it is apparent that applying Bose-Einstein statistics to ``phonons" obtained from the velocity distribution of particles is inappropriate for fluids.
The decomposition in Eq.~(\ref{eq:internal-energy-3}) is not valid for liquids. Only the total energy $E_{\rm tot}$ has a physical reality.

% -------------------------------------------------
\subsection{Sodium}
\label{sec:Na}
The next example is metallic Na, having a BCC structure in the solid state.
\begin{figure}[htbp]
  \centering
     \includegraphics[width=120 mm, bb=0 0 418 480]{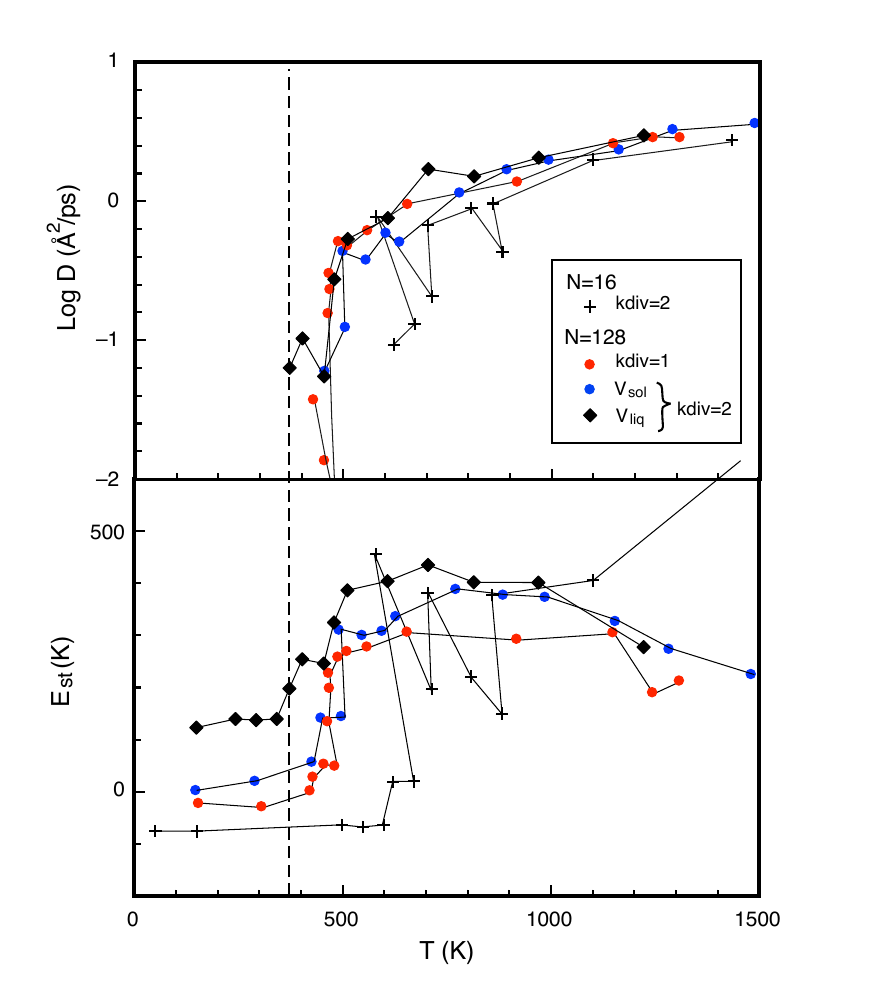} 
  \caption{The diffusion coefficient, $D$, and structural energy, $E_{\rm st}$, of Na calculated by NC+GGA. Cell sizes are $N=16$ and 128. The $k$ meshes are $2^{3}$ for $N=16$, whereas are $1$ (red circles) and $2^{3}$ (blue circles and black diamonds) for $N=128$. All the data except marked by diamond were obtained using the cell volume of crystal ($V_{\rm sol}$). Diamond data indicate using the liquid volume ($V_{\rm liq}$). $E_{\rm cut}=245$ eV. The experimental value for $T_{m}$ is indicated by the dashed line. For each $N$, the energy origin was set to the ground state energy for that cell. 
  } \label{fig:Est-lnD-Na}
\end{figure}
Cell sizes of 16 and 128 atoms were examined. In Fig.~\ref{fig:Est-lnD-Na}, the structural energy, $E_{\rm st}$, and diffusion constant, $D$, are plotted as functions of $T$. In this case, even though a finite width of the transition region is observed, there is no region of negative specific heat.
Let us first check the cell size dependence on $T_{m}$. When different sizes of supercells are compared, the equivalent $k$ meshes should be employed. Thus, the data obtained by a $2^{3}$ $k$-mesh of 16-atom cell (crosses in the figure) should be compared with the data obtained by a $1$ $k$-mesh of 128-atom (red circles). Clearly, the $T_{m}$ value is lowered from 600 to 460 K by increasing the cell size from 16 to 128. Accordingly, we can expect further decrease in the $T_{m}$ value, approaching the experimental value 371 K, by increasing the cell size.

For metallic systems, the $k$ mesh can affect the total energy calculation largely. Hence, the dependence of the $k$ mesh was also examined for the 128-atom cell. However, there was only a marginal change in $T_{m}$ between $1$ (red circles) and $2^{3}$ (blue circles) meshes, as shown in this figure.
Finally, the effect of cell volume was examined. The volume of the liquid phase is expanded from that of the solid by 9.4 \%. The date marked by diamonds in the figure indicate the results of the melting simulations on the expanded unit cells.
The effect of volume expansion is apparent. If we determine $T_{m}$ by the onset of the diffusion coefficient, we already obtain a good agreement with the experimental value, $T_{m}=371$ K.
The agreement in $T_{m}$ was also reported by Raty {\it et al.}~\cite{Raty07} for the pressure dependence of $T_{m}$ of Na: they used $N=128$. They reported that the agreement was obtained without use of the two-phase coexistence method; however, the agreement is not clear at $p=0$. This agreement in $T_{m}$ further confirms that the presence of surface/interface is not a universal requirement for calculation of $T_{m}$ for bulk materials.

There is a view that the equilibrium concentration of lattice vacancies reduces the melting temperature of bulk crystals through an increase in the melting entropy. The overestimation of $T_{m}$ by MD simulation is sometimes ascribed to this effect. This is tantamount to say that the observed $T_{m}$ is not the property of the perfect crystal. From the historical perspective, the effect of vacancies near $T_{m}$ was proposed for sodium by observing the correlation between the increases in the specific heat and thermal expansivity near $T_{m}$ \cite{Martin67,Sullivan64}. The formation energy of vacancy was estimated as 0.12-0.16 eV, and hence non-negligible concentrations of vacancies are certainly present at $T_{m}$. However, today, it is recognized that these increases are better described by the anharmonic phonon effects.
This vacancy mechanism is often repeated to correct the overestimation of $T_{m}$. For example, Chaplot introduced 1 \% vacancies for MgSiO$_{3}$ to reduce the calculated $T_{m}$ \cite{Chaplot98}: however, see a criticism by Belonoshko \cite{Belonoshko01}. For MgO, Arkhipin {\it et al.}~introduced about 20 \% of voids to reduce $T_{m}$ by 500 K \cite{Arkhipin24}. Such high concentrations of vacancy are unrealistic, unless the formation energy of defects is very small.

% -------------------------------------------------
\subsection{Silica}
\label{sec:silica}
\begin{figure}[htbp]
  \centering
    \includegraphics[width=120 mm, bb=0 0 425 500]{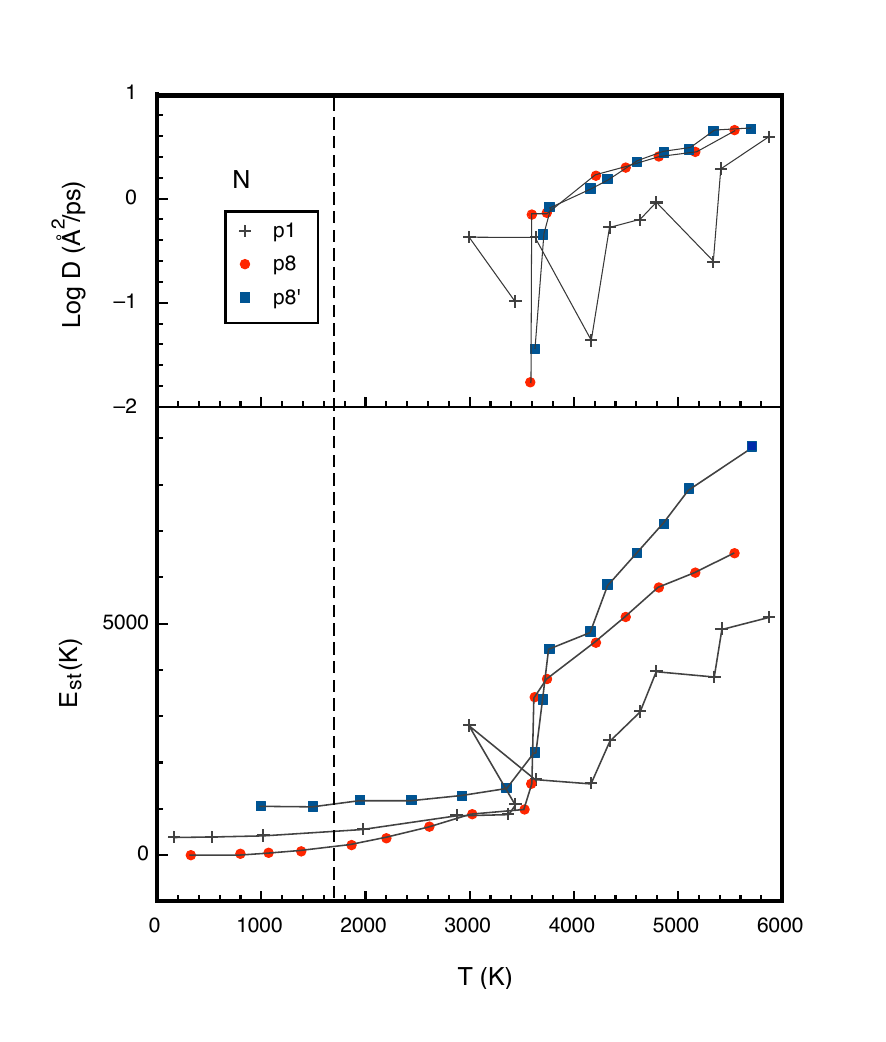} 
  \caption{The diffusion coefficient, $D$, and structural energy, $E_{\rm st}$, for $\alpha$-quartz calculated by NC+LDA. $E_{\rm cut}=544$ eV. p1 indicates results obtained with a $1 \times 1 \times 1$ cell ($4^{3}$ $k$-mesh), while p8 indicates a $2 \times 2 \times 2$ cell ($1$). p8' indicates the same cell with a volume expansion of 16.9 \%. The experimental value of $T_{m}$ is indicated by the dashed line. For each $N$, the energy origin was set to the ground state energy of that cell. For p8', it was set to that for the p8 cell.
  } \label{fig:Est-lnD-silica}
\end{figure}

The last example is an oxide material, namely, crystalline $\alpha$-quartz. The authors previously used this crystal as the parent material for the formation of silica glass and found the overestimation of $T_{m}$ \cite{Shirai22-Silica}. The primitive unit cell of $\alpha$-quartz has the rhombohedral Bravais lattice comprising three SiO$_{2}$ units. The data for $\alpha$-quartz listed in Table \ref{tab:sum-Tm} represent the results of MD simulations performed by the Phase code.  
Unfortunately, both codes Phase and VASP failed to keep long time MD simulations with a smaller cell having dimension of $1 \times 1 \times 1$. Drastic error propagations in atom forces occurred at some steps, leading to a break in the simulation. Osaka2k code using NC-LDA was found to be stable against the long run of MD simulations using this cell size.\cite{note-breakMD} Hence, the cell-size dependence for $\alpha$-quartz was examined by osaka2k code with the NC-LDA potential, using $E_{\rm cut}$=544 eV for the planewave expansion. The results are shown in Fig.~\ref{fig:Est-lnD-silica}. Although this cutoff energy is insufficient to achieve convergence of the total energy for oxide materials, it is meaningful for the purpose of examining the size dependence. As the result of this low cutoff energy, the potential of oxygen becomes effectively softer, which makes it easier to test the size dependence. The calculated value $T_{m}$ is reduced to 3500 K, as shown in Fig.~\ref{fig:Est-lnD-silica}, compared with that of 4500 K in Table \ref{tab:sum-Tm}. Even so, the value of $T_{m}=3500$ K is still much higher than the experimental result of 1700 K.

As shown in Fig.~\ref{fig:Est-lnD-silica}, the onset of melting almost does not change upon increasing the cell size from 9 to 72 atoms. The calculated $T_{m}$ seems rather to be saturated at the large overestimated value. Similar results were obtained for $\beta$-cristobalite in that increasing the unit cell dimension from $1 \times 1 \times 1$ to $2 \times 2 \times 2$ did not alter the overestimated $T_{m}$ (see Supplemental Materials). Generally, the size dependence is significant for small $N$ and is reduced as $N$ increases: for example, see Figs.~5-7 of model potentials by Hong and van de Walle \cite{Hong13}. The behavior of almost no change in $T_{m}$ between p1 and p8 implies that no significant reduction is likely to occur by further increase of $N$.
Furthermore, the effect of volume expansion was examined. The actual volume of the liquid silica is larger than that of the crystalline $\alpha$-quartz by 16.9 \%. The expanded cell by this amount was used, whose results are indicated by the p8' data in Fig.~\ref{fig:Est-lnD-silica}. Even with this large volume expansion, there is no discernible change in $T_{m}$. This is a contrasting behavior to the Na case. This insensitivity to the volume expansion suggests that there is some strong cause to fix the overestimation of $T_{m}$, irrespective of the cell volume as well as $N$.

We consider that the primal cause of the overestimation of $T_{m}$ for $\alpha$-quartz is the error of the potential. The systematic error of LDA is well known in that LDA overestimates the binding energy and accordingly underestimates the lattice parameters about a few \% \cite{Jones89}. The LDA error is a result of adapting the energy functional obtained for homogeneous electron systems to localized systems. Hence, the error generally becomes larger for strongly localized states. The underestimation of lattice parameters in LDA is fixed by GGA \cite{Jones89,PBE96,Zhang98,Perdew03}. 
In fact, in the calculation of $T_{m}$ of MgO by the coexistence method, Alf\'{e} showed a reduction of $T_{m}$ by GGA compared to LDA \cite{Alfe05} \cite{note-GGA-LDA}: the amount of the reduction is about 500 K.
His result already indicates that the calculated $T_{m}$ value varies depending on the potential even within DFT. The reduction in $T_{m}$ by GGA, however, does not guarantee that GGA is good enough. The overestimate of the binding energy still remains in GGA. Oxygen atom has the largest error. Patton {\it et al}.~report that the overbinding $\Delta$ by GGA is 0.53 eV for O$_{2}$ molecule \cite{Patton97}. For H$_{2}$O, $\Delta$ is only 0.03 eV, but this amounts to a change in $T_{m}$ by 300 K. When calculating the formation energies of metal oxides, it is a common practice to adjust the oxygen contribution by setting approximately 0.7 eV/(O atom) \cite{Wang06-GGA+U,Jain11,Mutter20}. Presently, there is no way to cope this adjustment in the MD simulation of melting.

The cohesive energy, $E_{\rm coh}$, is the energy difference between an isolated atom and the solid state, representing the extrema of localized and extended states, respectively. Therefore, the LDA/GGA error appears in $E_{\rm coh}$ most significantly. Because $H_{m}$ is the energy difference between the solid and liquid states, a similar effect is expected. Here is, however, a subtle problem of the electron correlation in liquids. Normally, the electron system having a high atom density represents the extended state (solid state), while an electron system having a low atom density represents the localized state (atom state). However, the average density of a liquid is close to that of the corresponding solid: the difference is only a few \%. There are even cases, such as Si, in which the liquid state is denser than the solid state. 
Therefore, the average density and average atom distance are not suitable to estimate the degree of localization of electrons for the case of liquid. Rather, in Sec.~\ref{sec:Sim-melt}, we emphasized the importance of the time correlation in atom positions.

The energy dissipation can be described by the time correlation in a given pair of atoms (\cite{Egelstaff-2ed}, Chap.~9),
\begin{equation}
G_{ij}({\bf r}, t) = \overline{ \delta({\bf r} +{\bf R}_{i}(t')-{\bf R}_{j}(t'+t)) }.
\label{eq:correlation_ij}
\end{equation}
The bar indicates the time average with respect to $t'$. The distance between those atoms in a particular atom pair, $R_{ij}$, immediately increases with time after melting. This correlation function may be approximately presented by $G_{ij}({\bf r}, t) \approx e^{-t/\tau_{ij}} \delta({\bf r}-\bar{d})$, where $\bar{d}$ is the average atom distance. The first factor $e^{-t/\tau_{ij}}$ is important. The relaxation time $\tau_{ij}$ is determined by the energy barrier between atoms $i$ and $j$. This corresponds to the saddle point of the energy landscape.
In contrast, RDF is obtained differently. The correlation is evaluated first by taking the particle average and then by taking the time average, as
\begin{equation}
\bar{G}({\bf r}, t) = \overline{ \langle \delta({\bf r} +{\bf R}_{i}(t')-{\bf R}_{j}(t'+t)) \rangle_{i,j} }.
\label{eq:ave-correlation}
\end{equation}
This correlation reflects only the average atom distance, which corresponds to the minima in the energy landscape. The information of energy dissipation is lost, and thus the diffusing nature of liquid is masked in RDF. Agreement not only in the energy minima but also in the saddle points is needed to reproduce the thermodynamic properties of liquids. Consequently, even though the averaged density does not change significantly on melting, it is likely that a significant change occurs in the electron correlation. Thus, the LDA/GGA error also affects the energy barrier between the solid and liquid states, leading to an overestimation of $T_{m}$.

Lastly, it is worthy to mention a distinct feature of the liquid phase of silica. Normally, significant changes in $E_{\rm st}$ occur only at phase transitions. However, as shown in Fig.~\ref{fig:Est-lnD-silica}, for the silica liquid, an apparent increase in $E_{\rm st}$ with increasing $T$ persists over the entire temperature range examined. This implies that the silica liquid is comprised of numerous distinct metastable states, which are separated each other by energy barriers. This could explain why the silica liquid has the highest viscosity among various liquids. Even in the liquid of silica, hysteresis is observed in the specific heat measurement \cite{Richet84a}. However, many things about the thermodynamic properties of silica liquid still remain unknown.

%%%%%%%%%%%%%%%%%%%%%%%%%%%%%%%%%%%%%%%%%%
\section{Conclusion}
\label{sec:Conclusion}
The difficulty in calculating $T_{m}$ consists in the thermodynamic nature of liquids. The thermodynamic temperature is determined by the time average of the kinetic energy, through Eq.~(\ref{eq:equil-temp}). This average is influenced by the atom coordinates at equilibrium. For solids, the equilibrium atom positions are unique and $T$ is determined by the bottom part of potential only. The equilibrium state is insensitive to the relaxation processes in which that state is led. 
In contrast, for liquids, the atom coordinate to determine the equilibrium continuously changes, and the temperature is determined by transitions between different coordinates. The details of energy relaxation processes thus affects $T_{m}$. The relaxation is controlled by the energy barriers, which are saddle points in the potential. One implication of this is that this explains why two similar potentials that give the almost same thermodynamic properties yield very different values for $T_{m}$, unless the two potentials are constructed so as to match up to the saddle points. Another one is that introduction of artificial heat bath should be avoided, because it alters the native balance among different relaxations. Adiabatic MDs are recommended.

The present simulations based on the adiabatic FP-MD resolved the presence of the transition width $W_{m}$, which is unavoidable for finite-size cells. The width can be removed only when the thermodynamic limit, $N \rightarrow \infty$, is taken. In particular, this width becomes harmful when the transition region exhibits negative specific heat. The extrema in the transition region could produce spurious equilibria. Bearing these cautions in mind, we have deduced the following conclusions from the present simulations.
First, a good agreement in $T_{m}$ is obtained for Si and Na by the adiabatic FP-MD simulations up to the uncertainty $W_{m}$. The fact that these agreements are obtained by the modest size of cells, approximately 100 atoms, encourages us to use the standard method of FP-MD simulations. This also implies that the involvement of surface in determining $T_{m}$ for bulk solids cannot be a universal reasoning of use of the two-phase coexistence method, unless the surface energy or other factors are too large.
Second, there are many cases in which the adiabatic FP-MD simulations give large overestimations for $T_{m}$. These cases are mainly those materials containing oxygen atoms. It is highly likely that the overestimation in $T_{m}$ is brought about by the LDA/GGA error in the electron correlation functional. The strong insensitivity against the change in $N$ and in the cell volume supports this conclusion.

\ack
The authors thank Prof.~Trachenko for discussing the specific heat of glasses and liquids.
The authors thank FORTE Science Communications (https://www.forte-science.co.jp/) for English language editing.

\section*{References}

\providecommand{\newblock}{}


\begin{thebibliography}{100}
\expandafter\ifx\csname url\endcsname\relax
  \def\url#1{{\tt #1}}\fi
\expandafter\ifx\csname urlprefix\endcsname\relax\def\urlprefix{URL }\fi
\providecommand{\eprint}[2][]{\url{#2}}

\bibitem{Morris94}
Morris J~R, Wang C~Z, Ho K~M and Chan C~T 1994 {\em Phys. Rev. B\/} {\bf 49}
  3109

\bibitem{Belonoshko94}
Belonoshko A~B 1994 {\em Geochim. Cosmochim. Acta\/} {\bf 58} 4039

\bibitem{Alfe05}
Alf\'{e} D 2005 {\em Phys. Rev. Lett.\/} {\bf 94} 235701

\bibitem{Bruesch82-1}
Bruesch P 1982 {\em Phonons: Theory and Experiments I ---Lattice dynamics and
  models of interatomic forces\/} (Berlin: Springer)

\bibitem{Porter97}
Porter L, Justo J~F and Yip S 1997 {\em J. Appl. Phys.\/} {\bf 82} 5378

\bibitem{Grimvall74}
Grimvall G and Sj\"{o}din S 1974 {\em Phys. Scr.\/} {\bf 10} 340

\bibitem{Granato10}
Granato A~V, Joncich D~M and Khonik V~A 2010 {\em Appl. Phys. Lett.\/} {\bf 97}
  171911

\bibitem{Yamahara01}
Yamahara K, Okazaki K and Kawamura K 2001 {\em J. Non-Cryst. Solids\/} {\bf
  291} 32

\bibitem{Vollmayr96}
Vollmayr K, Kob W and Binder K 1996 {\em Phys. Rev. B\/} {\bf 54} 15808

\bibitem{Kuzuu04}
Kuzuu N, Yoshie H, Tamai Y and Wang C 2004 {\em J. Non-Cryst. Solids\/} {\bf
  349} 319

\bibitem{Takada04}
Takada A, Richet P, Catlow C~R~A and Price G~D 2004 {\em J. Non-Cryst.
  Solids\/} {\bf 345-346} 224

\bibitem{Carre08}
Carr\'{e} A, Horbach J, Ispas S and Kob W 2008 {\em EPL\/} {\bf 82} 17001

\bibitem{Geske16}
Geske J, Drossel B and Vogel M 2016 {\em AIP Advances\/} {\bf 6} 035131

\bibitem{Niu18}
Niu H, Piaggi P~M, Invernizzi M and Parrinello M 2018 {\em Proc. Natl. Acad.
  Sci.\/} {\bf 115} 5348

\bibitem{Castleton09}
Castleton C~W~M, H\"{o}glund A and Mirbt S 2009 {\em Modelling Simul. Mater.
  Sci. Eng.\/} {\bf 17} 084003

\bibitem{Freysoldt14}
Freysoldt C, Grabowski B, Hickel T and Neugebauer J 2014 {\em Rev. Mod.
  Phys.\/} {\bf 86} 253

\bibitem{Hong13}
Hong Q and van~de Walle A 2013 {\em J. Chem. Phys.\/} {\bf 139} 094114

\bibitem{Jones89}
Jones R~O and Gunnarsson O 1989 {\em Rev. Mod. Phys.\/} {\bf 61} 689

\bibitem{Parr-Yang89}
Parr R~G and Yang W 1989 {\em Density-Functional Theory of Atoms and
  Molecules\/} (Oxford: Oxford)

\bibitem{note-checkDFT-liquids}
In current DFT studies on liquids, the calculated and experimental RDF are
  often compared in order to assess the accuracy of the calculations
  \cite{Stich91,Kresse94, Chelikowsky01,Massobrio15}. However, as noted above
  when discussing derivation of the model potential for silica, the RDF is
  insensitive for describing liquid properties. The RDF reflects the minima in
  the potential whereas energy dissipation is determined from the saddle points
  of the energy landscape, which are not included in the RDF.

\bibitem{Granato02}
Granato A~V 2002 {\em J. Non-Cryst. Solids\/} {\bf 307-310} 376

\bibitem{Trachenko11}
Trachenko K and Brazhkin V~V 2011 {\em Phys. Rev. B\/} {\bf 83} 014201

\bibitem{Bolmatov12}
Bolmatov D, Brazhkin V~V and Trachenko K 2012 {\em Sci. Rep.\/} {\bf 2} 421

\bibitem{Trachenko16}
Trachenko K and Brazhkin V~V 2016 {\em Rep. Prog. Phys.\/} {\bf 79} 016502

\bibitem{Proctor20}
Proctor J~E 2020 {\em Phys. Fluids\/} {\bf 32} 107105

\bibitem{Baggioli21}
Baggioli M and Zaccone A 2021 {\em Phys. Rev. E\/} {\bf 104} 014103

\bibitem{Shirai22-SH}
Shirai K, Watanabe K and Momida H 2022 {\em J. Phys.: Condens. Matter\/} {\bf
  34} 375902

\bibitem{Shirai22-Silica}
Shirai K, Watanabe K, Momida H and Hyun S 2023 {\em J. Phys.: Condens.
  Matter\/} {\bf 35} 505401

\bibitem{Mei92}
Mei J and Davenport J~W 1992 {\em Phys. Rev. B\/} {\bf 46} 21

\bibitem{Sugino95}
Sugino O and Car R 1995 {\em Phys. Rev. Lett.\/} {\bf 74} 1823

\bibitem{Frenkel96}
Frenkel D and Smit B 1996 {\em Understanding Molecular Simulation\/} (San
  Diego: Academic)

\bibitem{Cheng19}
Cheng B, Engel E, Behlerb J, Dellagod C and Ceriotti M 2019 {\em Proc. Nat.
  Acad. Sci.\/} {\bf 116} 1110

\bibitem{Belonoshko01}
Belonoshko A~B 2001 {\em Am. Mineral.\/} {\bf 86} 193

\bibitem{Usui10}
Usui Y and Tsuchiya T 2010 {\em J. Earth Sci.\/} {\bf 21} 801

\bibitem{Hong15}
Hong Q and van~de Walle A 2015 {\em Phys. Rev. B\/} {\bf 92} 020104(R)

\bibitem{Geng24}
Geng M and Mohn C 2024 {\em Phys. Rev. B\/} {\bf 109} 024106

\bibitem{Belonoshko06}
Belonoshko A~B, Skorodumova N~V, Rosengren A and Johanesson B 2006 {\em Phys.
  Rev. B\/} {\bf 73} 012201

\bibitem{Alfe11}
Alf\'{e} D, Cazoria C and Gillian M~J 2011 {\em J. Chem. Phys.\/} {\bf 135}
  024102

\bibitem{Cahn86}
Cahn R~W 1986 {\em Nature\/} {\bf 323} 668

\bibitem{Buffat76}
Buffat P and Borel J~P 1976 {\em Phys. Rev. A\/} {\bf 13} 2287

\bibitem{Lu98}
Lu K and Li Y 1998 {\em Phys. Rev. Lett.\/} {\bf 80} 4474

\bibitem{Jin01}
Jin Z~H, Gumbsch P, Lu K and Ma E 2001 {\em Phys. Rev. Lett.\/} {\bf 87} 055703

\bibitem{Bai05}
Bai X~M and Li M 2005 {\em J. Chem. Phys.\/} {\bf 123} 151102

\bibitem{Porter-PT-metals}
Porter D~A, Easterling K~E and Sherif M~Y 2009 {\em Phase Transformations in
  Metals and Alloys\/} 3rd ed (Boca Raton: CRC Press)

\bibitem{BornHuang}
Born M and Huang K 1969 {\em Dynamical Theory of Crystal Lattices\/} (Oxford:
  Clarendon)

\bibitem{Payne92}
Payne M~C, Teter M~P, Allan D~C, Arias T~A and Joannopoulos J~D 1992 {\em Rev.
  Mod. Phys.\/} {\bf 64} 1045

\bibitem{Egelstaff-2ed}
Egelstaff P~A 1992 {\em An Introduction to the Liquid State\/} 2nd ed (Oxford:
  Oxford)

\bibitem{Copey74}
Copley J~R~D and Rowe J~M 1974 {\em Phys. Rev. Lett.\/} {\bf 32} 49

\bibitem{Smith17}
Smith H, Li C~W, Hoff A, Garrett G~R, Kim D~S, Yang F~C, Lucas M~S, Swan-Wood
  T, Lin J~Y~Y, Stone M~B, Avernathy D~L, Demetriou M~D and Fultz B 2017 {\em
  Nature physics\/} {\bf 13} 900

\bibitem{Suck05}
Suck J~B 2005 Experimental investigations of collective excitations in
  disordered matter {\em Collective Dynamics of Nonlinear and Disordered
  Systems\/} ed Radons G, Just W and H\"{a}ussler P (Berlin: Springer) p 147

\bibitem{Wallace97b}
Wallace D~C 1997 {\em Phys. Rev. E\/} {\bf 56} 4179

\bibitem{Wallace98}
Wallace D~C 1998 {\em Phys. Rev. E\/} {\bf 57} 1717

\bibitem{Wallace02}
Wallace D~C 2002 {\em Statistical Physics of Crystals and Liquids: A guide to
  highly accurate equations of state\/} (Singapore: World Scientific)

\bibitem{Shu80}
Shu H~C, Gaur U and Wunderlich B 1980 {\em J. Polym. Sci.: Polym. Phys. Ed.\/}
  {\bf 18} 449

\bibitem{Leibfried61}
Leibfried G and Ludwig W 1961 Theory of anharmonic effects in crystals {\em
  Solid State Physics\/} vol~12 ed Seitz F and Turnbull D (New York: Adacemic)
  p 275

\bibitem{Perepezko84}
Perepezko J~H and Paik J~S 1984 {\em J. Non-Cryst. Solids\/} {\bf 61/62}
  113--118

\bibitem{Rhim00a}
Rhim W~K and Ishikawa T 2000 {\em Int. J. Thermophysics\/} {\bf 21} 429

\bibitem{Li04}
Li Q, Zhu Y~Y, Liu R~P, Li G, Ma M~Z, Yu J~K, He J~L, Tian Y~J and Wang W~K
  2004 {\em Appl. Phys. Lett.\/} {\bf 85} 558

\bibitem{note-phonon-liquids}
The term phonon is used here to mean oscillatory motions around the {\em
  equilibrium} position, the concept of which is valid only for solids. On the
  other hand, sound waves have reality for fluids too. However, there is no
  proof that sound waves obey the Bose-Einstein statistics. In fact, for the
  extreme case of ideal gases, sound waves are not needed to describe the
  specific heat. The Bose-Einstein and Fermi-Dirac statistics can be applied
  only to independent quasiparticle systems \cite{Reif}.

\bibitem{Alvarez-Donado20}
Alvarez-Donado R and Antonelli A 2020 {\em Phys. Rev. Research\/} {\bf 2}
  013202

\bibitem{Han20}
Han D, Wei D, Yang J, Li H~L, Jiang M~Q, Wang Y~J, Dai L~H and Zaccone A 2020
  {\em Phys. Rev. B\/} {\bf 101} 014113

\bibitem{Birge85}
Birge N~O and Nagel S~R 1985 {\em Phys. Rev. Lett.\/} {\bf 54} 2674

\bibitem{Nielsen96}
Nielsen J~K and Dyre J~C 1996 {\em Phys. Rev. B\/} {\bf 54} 15754

\bibitem{Hentschel08}
Hentschel H~G~E, Ilyin V, Procaccia I and Schupper N 2008 {\em Phys. Rev. E\/}
  {\bf 78} 061504

\bibitem{note-Relax}
The importance of relaxation in calculating the melting temperature can be seen
  in comparison of the calculated $T_{m}$ between different MD methods. Using
  the same potential used in the thermodynamic integration method \cite{Mei92},
  Morris {\it et al}.~showed that the obtained $T_{m}$ was reduced by about 80
  K in the coexistence method \cite{Morris94}. The details of relaxation
  process are affected by changing MD methods to this extent.

\bibitem{Davies53}
Davies R~O and Jones G~O 1953 {\em Proc. Roy. Soc. A\/} {\bf 217} 26

\bibitem{Davies53a}
Davies R~O and Jones G~O 1953 {\em Adv. Phys.\/} {\bf 2} 370

\bibitem{Shirai24-hysteresis}
Shirai K, arXiv:2406.15726

\bibitem{Shirai20-GlassState}
Shirai K 2020 {\em J. Phys. Commun.\/} {\bf 4} 085015

\bibitem{PHASE}
Yamasaki T, Kuroda A, Kato T, Nara J, Koga J, Uda T, Minami K and Ohno T 2019
  {\em Comp. Phys. Commun.\/} {\bf 244} 264

\bibitem{VASP}
G. Kresse and J. Furthmu\"{u}ller, Software VASP, Vienna, (1999); G. Kresse and
  J. Furthmu\"{u}ller, Phys Rev B {\bf 54} 11169 (1996)

\bibitem{Martin04}
Martin R~M 2004 {\em Electronic Structure: Basic theory and practical
  methods\/} (Cambridge: Cambridge)

\bibitem{Paola16}
Paola C~D and Brodholt J~P 2016 {\em Sci. Rep.\/} {\bf 6} 29830

\bibitem{Hernandez22}
Hernandez J~A, Mohn C~E, Guren M~G, Baron M~A and Tronnes R~G 2022 {\em
  Geophys. Res. Lett.\/} {\bf 49} e2021GL097262

\bibitem{Cheng24}
Cheng M and Sun H 2024 {\em Phys. Rev. Mater.\/} {\bf 8} 113604

\bibitem{CRC92}
 2011 {\em CRC Handbook of Chemistry and Physics\/} ed Haynes W~M (Boca Raton:
  CRC Press) pp 6--146 92nd ed

\bibitem{Richet82}
Richet P, Bottinga Y, Denielou L, Pettitet J~P and T\'{e}qui C 1982 {\em
  Geochim. Cosmochim. Acta\/} {\bf 46} 2639

\bibitem{Kittel88}
Kittel C 1988 {\em Phys. Today\/} {\bf 41} (5) 93

\bibitem{Mandelbrot89}
Mandelbrot B 1989 {\em Phys. Today\/} {\bf 42} (1) 71

\bibitem{Lynden-Bell99}
Lynden-Bell D 1999 {\em Physica A\/} {\bf 263} 293--304

\bibitem{Landsberg87}
Landsberg P~T and Peari\'{c} J~E 1987 {\em Phys. Rev. A\/} {\bf 35} 4397

\bibitem{Michaelian07}
Michaelian K and Santamaria-Holek I 2007 {\em EPL\/} {\bf 79} 43001

\bibitem{Schmidt21}
Schmidt M, Kusche R, Hippler T, Donges J, Kronm\"{u}ller W, von Issendorff B
  and Haberland H 2001 {\em Phys. Rev. Lett.\/} {\bf 86} 1191

\bibitem{BHS82}
Bachelet G~B, Hamann D~R and Schl\"{u}ter M 1982 {\em Phys. Rev. B\/} {\bf 26}
  4199

\bibitem{TM91}
Troullier N and Martins J~L 1991 {\em Phys. Rev. B\/} {\bf 43} 1993

\bibitem{Alfe03}
Alf\'{e} D and Gillan M~J 2003 {\em Phys. Rev. B\/} {\bf 68} 205212

\bibitem{Shirai-EntropyLiquid}
Shirai K, Momida H, Sato K, and Hyun S, arXiv:2411.10930

\bibitem{Raty07}
Raty J~Y, Schwegler E and Bonev S~A 2007 {\em Nature\/} {\bf 449} 448

\bibitem{Martin67}
Martin D~I 1967 {\em Phys. Rev.\/} {\bf 154} 571

\bibitem{Sullivan64}
Sullivan G~A and Weymouth J~W A1141 {\em Phys. Rev.\/} {\bf 136} 024106

\bibitem{Chaplot98}
Chaplot S~L, Choudhury N and Rao K~R 1998 {\em Am. Mineral.\/} {\bf 83} 937

\bibitem{Arkhipin24}
Arkhipin A~S, Pisch A, Uspenskaya I~A and Jakse N 2024 {\em Ceramics\/} {\bf 7}
  1187

\bibitem{note-breakMD}
The problem of Phase and VASP codes could have been related to core-charge
  correction employing in ultrasoft potential and PAW methods. The correcting
  charge varies largely in the core region. It is likely that the numerical
  errors in the atom forces due to this core-charge correction coupled with the
  large energy fluctuation of Eq.~(12) due to the small cell size, led to the
  error propagation in atom forces.

\bibitem{PBE96}
Perdew J~P, Burke K and Ernzerhof M 1996 {\em Phys. Rev. Lett.\/} {\bf 77} 3865

\bibitem{Zhang98}
Zhang Y and Yang W 1998 {\em Phys. Rev. Lett.\/} {\bf 80} 890

\bibitem{Perdew03}
Perdew J~P and Kurth S 2003 Density functionals for non-relativistic coulomb
  systems in the new century {\em A Primer in Density Functional Theory\/} ed
  Fiolhais C, Nogueira N and Marques M (Berlin: Springer)

\bibitem{note-GGA-LDA}
Alf\'{e} warned, in \cite{Alfe05}, not to take this---GGA improves the LDA
  overestimation---as a general trend, because of existence of the counter
  examples \cite{Sugino95,Alfe03}. However, these counter examples are those
  results of thermodynamic integration, for which the mechanism of error may be
  different from that of the coexistence method. See Sec.~3.2.

\bibitem{Patton97}
Patton D~C, Porezag D~V and Pederson M~R 1997 {\em Phys. Rev. B\/} {\bf 55}
  7454

\bibitem{Wang06-GGA+U}
Wang L, Maxisch T and Ceder G 2006 {\em Phys. Rev. B\/} {\bf 73} 195107

\bibitem{Jain11}
Jain A, Hautier G, Ong S~P, Moore C~J, Fisher C~C, Persson K~A and Ceder G 2011
  {\em Phys. Rev. B\/} {\bf 84} 045115

\bibitem{Mutter20}
Mutter D, Urban D~F and Els\"{a}sser C 2020 {\em Materials\/} {\bf 13} 4303

\bibitem{Richet84a}
Richet P 1984 {\em Geochim. Cosmochim. Acta\/} {\bf 48} 471

\bibitem{Stich91}
Stich I, Car R and Parrinello M 1991 {\em Phys. Rev. B\/} {\bf 44} 11092

\bibitem{Kresse94}
Kresse G and Hafner J 1994 {\em Phys Rev B\/} {\bf 49} 14251

\bibitem{Chelikowsky01}
Chelikowsky J~R, Derby J~J, Godlevsky C~V, Jain M and Raty J~Y 2001 {\em J.
  Phys.: Condens. Matter\/} {\bf 13} R817

\bibitem{Massobrio15}
Massobrio C, Du J, Bernasconi M and Salmon P~S 2015 {\em Molecular Dynamics
  Simulations of Disordered Materials: from network glasses to phase-change
  memory alloys\/} (Heidelberg: Springer)

\bibitem{Reif}
Reif F 1965 {\em Fundamentals of Statistical and Thermal Physics\/} (Tokyo:
  McGraw-Hill Kogakusha)

\end{thebibliography}
\end{document}